\documentclass[twocolumn,prl]{revtex4}

  \usepackage[dvips]{graphicx}

\usepackage{dcolumn}
\usepackage{amsmath}
\usepackage{latexsym}
\usepackage{amssymb}

\begin{document}


\title[Short Title]{Quantum phase slip interference device based on superconducting nanowire}
\author{T. T. Hongisto and A. B. Zorin}
\affiliation{Physikalisch-Technische Bundesanstalt, Bundesallee
100, 38116 Braunschweig, Germany}

\date{October 31, 2011}

\begin{abstract}

We propose a transistor-like circuit including two serially connected segments of a narrow
superconducting nanowire joint by a wider segment with a capacitively coupled gate in between.
This circuit is made of amorphous NbSi film and embedded in a network of on-chip Cr microresistors ensuring a
sufficiently high external electromagnetic impedance. Assuming a virtual regime of quantum phase slips (QPS)
in two narrow segments of the wire, leading to quantum interference of voltages on these segments, this circuit
is dual to the dc SQUID. Our samples demonstrated appreciable Coulomb blockade voltage (analog of critical
current of the SQUIDs) and periodic modulation of this blockade by an electrostatic gate (analog of flux
modulation in the SQUIDs). The model of this QPS transistor is discussed.


\verb  PACS numbers: 73.63.Nm, 73.23.Hk, 74.78.Na, 85.25.Am, 74.81.Fa

\end{abstract}
\maketitle

The discovery of the Josephson effect \cite{Josephson} triggered the development of superconductive
electronic devices operating on the principle of the classical behavior of the collective quantum variable,
i.e. the superconducting phase difference. These include rf and dc SQUIDs \cite{Clarke-Braginski},
single-flux-quantum logic circuits \cite{Likharev-Semenov}, Josephson voltage standard arrays \cite{Hamilton},
microwave receivers \cite{Likharev-Migulin}, etc.
The operation of these devices is based on the nonlinear (2$\pi$ periodic) dependence of the
supercurrent $I = I_c \sin \varphi$ on the phase difference $\varphi$ on the Josephson junction (JJ),
where $I_c$ is the critical current. The electromagnetic impedance of the JJ for
a small harmonic signal is therefore expressed in terms of Josephson inductance $L_J$
which varies with $\varphi$ also in a periodic fashion,
$L_J^{-1}(\varphi) = (2 \pi I_c/\Phi_0) \cos\varphi$, where $\Phi_0=h/2e$ is the flux quantum.

Recently, a new type of nonlinear superconducting element, dual to the JJ, -
the quantum phase slip (QPS) nanowire (see, for example, the review on the physical
properties of superconducting nanowires Ref.\,\cite{ArutyunovGolubevZaikin},
and references therein) -
was proposed in Refs.\,\cite{MooijHarmans,MooijNazarov}. Being included in a superconducting
ring with flux-bias near degeneracy points, $\Phi_b = (m+0.5)\Phi_0$, $m$ is an integer number,
the QPS element enables coupling of quantum states $|m\rangle$ and $|m+1\rangle$.
This leads to  anticrossing of corresponding levels with the gap $\Delta E= E_{\textrm{QPS}}$.
For a uniform nanowire of length $l$, the characteristic coupling energy $E_{\textrm{QPS}} = l \epsilon_{\textrm{S}}$
with $\epsilon_{\textrm{S}} \approx (k_B T_c/\xi)(R_Q/R' \xi)^{1/2} / \exp(0.3\,\alpha R_Q/R' \xi)$,
where $T_c$ is the critical temperature, $\xi$ the coherence length, $R'$ the normal-state resistance
per unit length, $R_Q = h/4e^2 \approx 6.5$\,k$\Omega$, the resistance quantum,
and $\alpha \sim 1$ \cite{ArutyunovGolubevZaikin,MooijHarmans}.
The two-level spectrum of this circuit  was proposed \cite{MooijHarmans} for using in
a flux qubit (compare with the dual charge qubit in which the coupling
of charge states is possible due to finite Josephson energy $E_J = (\Phi_0/2 \pi)I_c$ \cite{Nakamura}).

Coupling of the nanowire to a high-ohmic environment through the appropriate boundary
conditions imposes dramatic changes in its dynamics \cite{Buchler}.
Based on the quantum duality of the Josephson and QPS effects stemming from conjugation
of the phase and charge variables, Mooij and Nazarov had further
predicted \cite{MooijNazarov} that in the charge-bias regime, the QPS element should
demonstrate the Coulomb blockade behavior with
periodic dependence of voltage on the injected charge $q$,
viz. $V_{\textrm{QPS}}(q) = V_c \sin(\pi q/e)$ with $V_c = 2\pi E_{\textrm{QPS}}/2e$.
Thus, this element should behave as a nonlinear capacitor,
$C_{\textrm{QPS}}^{-1}(q) = dV_{\textrm{QPS}}/dq = (\pi V_c/e) \cos(\pi q/e)$ with
a 2$e$-periodic dependence on $q$.

\begin{figure}[b]
\begin{center}
\includegraphics[width=3.1in]{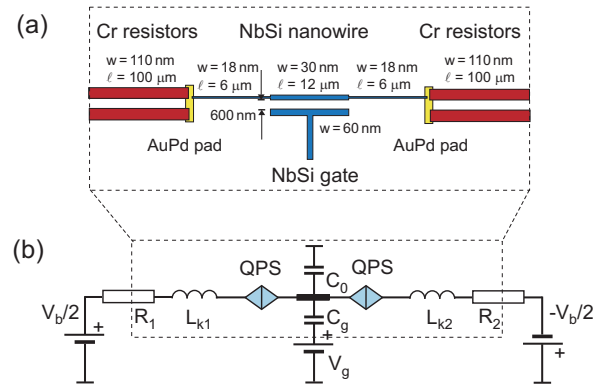}
\caption{(Color online) (a) The layout and (b) the simplified electric circuit diagram of the QPS transistor
embedded in the 4-terminal network of on-chip resistors.
The device includes two QPS elements denoted by diamond symbols, kinetic inductances of
the nanowire segments and a capacitive gate. Thicknesses of NbSi and Cr
films are 10 nm and 30 nm, respectively.} \label{fig:EqvSchm}
\end{center}
\end{figure}

It is clear that such a QPS element could be exploited in the engineering of electronic
devices controlled by the charge in the classical manner. These devices are dual to the Josephson ones
controlled by the classically behaved phase.
For example, applying microwave irradiation of frequency $f$ should lead to the formation of the
current steps in the $I$-$V$ curve at $I = 2nef$ ($n$ is integer), enabling a fundamental standard
of current \cite{MooijNazarov}.
These steps are dual to Shapiro voltage steps in a JJ $I$-$V$ curve at $V = n\Phi_0f$ \cite{Shapiro}.
Developing such an analogy, in this paper we proposed and realized a QPS-based single charge transistor which is dual
to dc SQUID and can be operated as an electrometer.

Our device (see schematic diagram in Fig.\,\ref{fig:EqvSchm}a) has a symmetric in-line configuration
and includes two narrow pieces of superconducting nanowire
joined by an island (a wider middle part of the same nanowire) with a capacitively coupled gate.
This transistor is embedded in the network of compact high-ohmic resistors,
$R \sim 0.4 $\,M$\Omega \gg R_Q$,
which ensure a sufficiently high electromagnetic impedance seen by the transistor. Due to this improvement,
a classical charge regime of operation with significant damping of dynamics is realized.
(Note that the recent proposal by Hriscu and Nazarov \cite{HriscuNazarov-trans}
deals with a complementary QPS transistor, without resistors, operating in a quantum regime,
which, in contrast to our circuit is phase biased and, therefore, not strictly dual to that of the
dc SQUID.)

The Kirchhoff equation for the equivalent electric circuit
presented in Fig.\,\ref{fig:EqvSchm}b has the form
\begin{equation}
V_1+V_2 = V_b, \;\, V_i = L_{ki}\ddot{q}_i + R_i \dot{q}_i + V_{\textrm{QPS}i}(q_i),\;  i = 1,\,2. \label{equation-of-motion}
\end{equation}
Here, $V_{\textrm{QPS}i}(q_i) = V_{ci} \sin(\pi q_i/e)$ are the $2e$-periodic voltages
on the QPS elements on either side of the island. The charge conservation
relation takes the form
\begin{equation}
q_1 - q_2 =-C_g V_g - (C_g+C_0) (V_1 - V_2).\label{q1q2-relation}
\end{equation}
This equation reflects the balance of the charges injected in the island
and the polarization charges induced by the gate and floating potential of the island having
a self-capacitance $C_0$ with respect to ground. In the case
of sufficiently small gate- and self-capacitances,
\begin{equation}
C_g, C_0 \ll e/V_{c1,c2}, \label{small-CgC0}
\end{equation}
and small injected currents (viz. $\pm \dot{q}_{1,2}$), the difference
charge $q_1-q_2 = -C_g V_g \equiv Q_g$ is totally controlled by the gate voltage $V_g$.

One can easily find that equations Eqs.~(\ref{equation-of-motion}-\ref{small-CgC0}) are
dual to those of the dc SQUID, see, for example, Ref.\,\cite{Likharev-book}. The charges $q_1$
and $q_2$ are dual to the phases across individual JJs, whereas
amplitudes $V_{c1,c2}$ play the roles of the critical currents. To continue this
analogy we can put in correspondence the pairs of parameters $L_{k1,k2}$ and JJ
capacitances, $R_{1,2}$ and shunting resistors, etc., for the QPS transistor and dc SQUID,
respectively. The condition of small gate- and self-capacitances Eq.~(\ref{small-CgC0})
is dual to the requirement of a small inductance of the SQUID loop $(L \ll L_J)$. Similar to external
flux control of quantum interference in the SQUID \cite{Likharev-book},
the voltage on the QPS transistor is resulted from quantum interference of voltages on
individual QPS elements and  controlled by the gate charge $Q_g$,
\begin{equation}
V = V_{\textrm{QPS}1} + V_{\textrm{QPS}2} = V_{m}(Q_g) \sin(\pi Q/e), \label{Vc-mod}
\end{equation}
where  $Q = q_1 + e\eta/\pi = q_2 + e\eta/\pi - Q_g$ is the "average" charge,
\begin{equation}
\tan \eta = \frac{2 V_{c2}\tan(\pi Q_g/2e)}{V_{c+} + V_{c-} \tan^2(\pi Q_g/2e)}, \label{eta-value}
\end{equation}
$V_{c\pm} = V_{c1}\pm V_{c2}$, and the Coulomb blockade voltage
\begin{equation}
V_m^2 = V_{c1}^2 + V_{c2}^2 + 2V_{c1} V_{c2} \cos(\pi Q_g/e). \label{Vm-ampl}
\end{equation}
In the case of a symmetrical circuit, $V_{c1}= V_{c2} = V_c$, $Q = q_{1,2} \pm Q_g/2$, Eq.~(\ref{Vm-ampl})
yields the maximum modulation of the blockade voltage, $V_m = 2V_c |\cos(\pi Q_g/2e)|.$
In the case of a highly asymmetrical circuit, say $V_{c2}/V_{c1} = a \ll 1$, charge $Q \approx q_1$
and the blockade voltage is only slightly modulated by the gate charge $Q_g$,
\begin{equation}
V_m = V_{c1} [1 + a \cos(\pi Q_g/e)]. \label{V-asymm-ampl}
\end{equation}
Thus the equation of motion Eq.\,(\ref{equation-of-motion}) takes a form of the
resistively shunted junction (RSJ) model for dc SQUID,
\begin{equation}
L_{k}\ddot{Q} + R \dot{Q} + V_m(Q_g)\sin(\pi Q/e) = V_b \label{equation-of-motion2}
\end{equation}
with $L_k = L_{k1} + L_{k2}$ and $R = R_2 + R_2$.

Equation~(\ref{equation-of-motion2}) describes the dynamics of a nonlinear oscillator with finite damping.
It yields a dc $IV$-curve with both the static- and running-charge (the oscillating
regime, $\omega = \pi\langle \dot{Q}\rangle/e = \pi\langle I\rangle/e$) branches, dependent on $Q_g$
in a periodic fashion. The dimensionless parameter, crucial for the dynamics,
introduced earlier in Ref.~\cite{MooijNazarov},
\begin{equation}
\beta_{\textrm{QPS}}= \omega_c L_k/R,  \label{beta-QPS}
\end{equation}
where $\omega_c = \pi V_m(Q_g)/eR$ is a characteristic circular frequency.
Eventually, $\beta_{\textrm{QPS}}$ is the analog of the Stewart-McCumber
parameter in the Josephson dynamics. In the most realistic case
of large damping, $\beta_{\textrm{QPS}} < 1$, the $IV$-curve has
a shape with characteristic back bending,
\begin{equation}
\langle V \rangle = [V_m^2(Q_g) + R^2\langle I\rangle^2]^{1/2}-R\langle I\rangle, \label{IV-shape}
\end{equation}
where $\langle V \rangle$ is the average voltage on the nanowire.

The fabrication method (some modification of the method developed in Ref.~\cite{vanderSar})
included the following steps.
The samples were fabricated on a silicon substrate, having 300\;nm thick thermal oxide layer,
combining two processes: shadow evaporation and the sputter deposition/etching step.
The Cr resistors were fabricated in a single vacuum cycle with the AuPd contact wires and
micropads utilizing the shadow evaporation technique through a bi-layer PMMA/Copolymer
stencil mask. The 30~nm thick Cr resistors were evaporated first at a low residual pressure of
oxygen ($\sim 10^{-6}$\,mbar) followed by a 50\;nm thick layer of AuPd from an angle for which the narrow stencil
openings for Cr resistors were overshadowed by the mask. Using this trick, the formation
of AuPd shadows parallel to the Cr resistors was avoided.

The Cr resistors and other parts of the circuit were protected by a PMMA mask while 10\;nm
thick amorphous Nb$_x$Si$_{1-x}$ film $(x = 0.45,$ defined
by calibrating the sputter rate for each element at a given sputter power and periodically
confirmed by EDX measurement using separately deposited films on Ge substrates) was co-sputtered on the substrate making
contact only with the AuPd micropads. These micropads allowed making
an electrical contact of the NbSi film with the Cr resistor circuitry.
To remove organic residue and water for the purpose of forming reliable contacts, we cleaned the
micropads in RIE with oxygen plasma and baked the sample overnight in N$_2$ atmosphere
at a temperature of 120\;C$^\circ$ before NbSi deposition. The subtrate was rotated 20 times
during a one-minute-long deposition. After the lift-off, the wafer was coated with
inorganic negative tone HSQ resist (XR-1541, Dow Corning) patterned with e-beam. After exposure,
the HSQ resist had characteristics of thermal silicon oxide and could be used as an etch mask to define the
NbSi nanowires along with the island and the gate structure made out of the same NbSi film.
In a final step, the ICP etch process with SF$_6$ gas was used for etching.

\begin{figure}
\begin{center}
\includegraphics[width=3.0in]{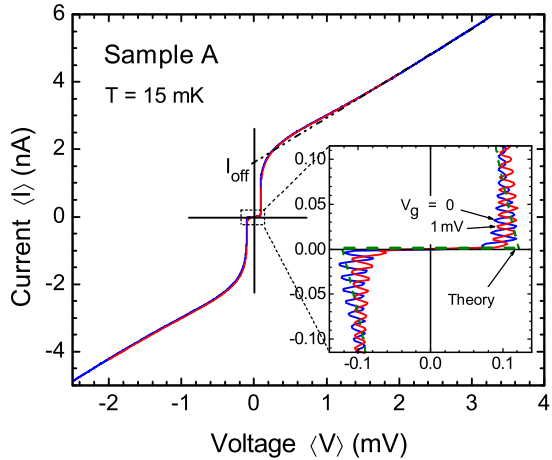}
\caption{(Color online) The charge-modulated $IV$-curves of Sample A recorded in a current bias regime
for two gate voltages shifted by a half-period. In the region of small currents (blown up in inset) one can see
the modulation with period $\Delta I = 13.5$\,pA, which is due to the asymmetry of off-chip
biasing circuitry resulting in the current
dependence of the electric potential of the transistor island, $\delta V = (R_{\textrm{bias}1}-R_{\textrm{bias}2})I$
and, therefore, of the effective gate charge, $\delta Q_g = (C_g + C_0)\, \delta V$
(cf Eq.~(\ref{q1q2-relation})). The green dashed line shows the shape of the
bare $IV$-curve given by the RSJ model Eq.~(\ref{IV-shape}).} \label{fig:IV-mod}
\end{center}
\end{figure}

\begin{figure}
\begin{center}
\includegraphics[width=3.2in]{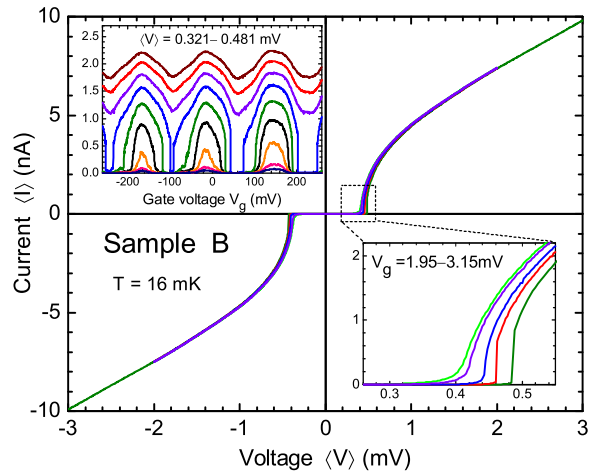}
\caption{(Color online) The $IV$-curves of Sample B measured in the
voltage bias regime at different
values of gate voltage $V_g$. The bottom right inset shows details of the Coulomb blockade
corner. Upper left inset: the gate voltage dependence of the transistor current measured at
different bias voltages $V_b$
providing a steady increase of $\langle V \rangle$ from 0.321\;mV up to 0.481\;mV in 4\;mV steps
(from bottom to top).} \label{IVs-largest-Vg-effect}
\end{center}
\end{figure}

The samples were measured in a dilution refrigerator with electrical lines
equipped with microwave frequency filters made of pieces of Thermocoax$^{\textrm{TM}}$ cable.
We used battery power sources and home-made electronics for either a voltage or a current bias of our circuits.
Most of the measurements were performed at the lowest temperature of the fridge, $T = 15$\,mK.
The layout allowed independent characterization of
both pairs of on-chip resistors. Their $IV$-curves at millikelvin temperatures were practically
linear with resistance about 20\% above the values measured at room temperature.
The NbSi films had the superconducting transition temperature $T_c \approx 1$\,K
and the normal state resistance per square about 550\,$\Omega$ (approaching the level of ca.
600\,$\Omega$ of the thickness-driven superconductor-to-insulator transition in somewhat
thicker films of Nb$_x$Si$_{1-x}$ with stoichiometry $x = 0.15$ \cite{Marrache-Kikuchi}), yielding for the narrow
segments the specific resistance $R' \approx 31$\,k$\Omega/\mu$m. The NbSi film parameters were
measured at $T=3.4$\,K on the stand-alone test wires 20\,nm in width, fabricated on the same chips.

Here we report on two samples (A and B, see their characteristics in Figs.\,2 and 3, respectively).
In comparison with Sample B whose dimensions are shown in Fig.\,1a,
Sample A had a twice longer island (25\,$\mu$m) and somewhat shorter QPS sections (4.5\,$\mu$m each).
The layout of the gate was identical in both samples.
Both samples showed an appreciable Coulomb blockade, $\langle I \rangle = 0$, and a gate effect at small bias.
An increase in current $\langle I \rangle \gtrsim 1$\,nA was accompanied by a steep rise in voltage $\langle V \rangle$
indicating a gradual turning of the wire into the normal state.
On a large scale, the $IV$-curves of both samples exhibited a positive excess current, i.e. had a shifted linear asymptote,
$I_{\textrm{off}} = \langle I \rangle - \langle V \rangle (d\langle I \rangle/d\langle V \rangle)_{\langle V \rangle \rightarrow \infty}
\approx 1.5-2.0$\;nA, indicating that part of the current at large bias was still
transported by a superconducting component \cite{Likharev-RMP79-review}.
This shape of the $IV$-curve argues that the observed gate effect has a superconductive origin
and allows us to rule out the strikingly different effect of the Coulomb blockade in a normal single electron
transistor, characterized by a pronounced voltage offset \cite{FultonDolan87} or, equivalently,
a negative "excess" current.

On a small scale, one can see the clear gate modulation of $IV$-curves, although with very different
periods, $\Delta V_g \approx 2$\,mV and 150\,mV, for Samples A and B, respectively.
Assuming the $2e$-periodic charge dependence in Sample A (see inset to Fig.\,2)
capacitance $C_g = 2e/\Delta V_g$ takes the value of about $160$\,aF.
This value is not far from the modelled value of $200$\;aF.
The depth of the modulation of about 40\% suggests - according
to our model Eq.\:(\ref{V-asymm-ampl}) - the ratio of QPS energies,
$a = V_{c2}/V_{c1} = E_{\textrm{QPS}2}/E_{\textrm{QPS}1} \approx 0.2$ whereas the average
value is about 20\,$\mu$eV. Inserting this value in the formula for the QPS energy and assuming
the length of the QPS junction of the order of the coherence length $\xi$, we obtain
the ratio $R_Q/R'\xi \approx$ 9 and, therefore, a rough estimate $\xi \approx $ 20\,nm.
This value is close to the values $\xi = 10-15$\,nm found in Ref.\,\cite{vanderSar}.

The shape of a bare (not modulated) $IV$-curve is only qualitatively similar
to that given by Eq.~(\ref{IV-shape}) and shown by the dashed line.
A back bending weaker than in theory can be attributed to the
effect of stray capacitance of the resistors, resulting in a roll-off of effective impedance $R$
in Eq.~(\ref{IV-shape}) with the rise of current ($\propto \omega$).
Rounding of the Coulomb blockade corner is attributed to
the effect of noise, omitted in our model.

Sample B demonstrated an even stronger gate effect (see upper left inset of
Fig.\;\ref{IVs-largest-Vg-effect}), but also the peculiar properties.
A huge period $\Delta V_g \approx 150$\;mV of the gate voltage dependence
corresponds to coupling capacitance $C_g \approx 2$\,aF.  This value is two orders smaller than
the designed value, so we conclude that the actual island in this sample had
much smaller dimensions (very crudely, on the order of 100\;nm) and located presumably
in one of the narrower segments of the nanowire. The available single gate did not allow any conclusions
about exact size and location of this island, whereas the depth of modulation of about 20\% yielded
the ratio of QPS energies,
$E_{\textrm{QPS}2}/E_{\textrm{QPS}1} \approx 0.1$ with $E_{\textrm{QPS}1} \approx 150\:\mu$eV.
A small hysteresis and almost vanishing back bent part in the $IV$-curve, seen in the lower
bottom inset of Fig.\;\ref{IVs-largest-Vg-effect}, may indicate that the value of the damping
parameter $\beta_{\textrm{QPS}}$ Eq.\,(\ref{beta-QPS}) is between 1 and 2 \cite{Likharev-book},
whereas our estimation yields a value of $\beta_{\textrm{QPS}}$ smaller by
an order of magnitude, ensuring a heavily overdamped regime.

Interestingly, the observed periodic pattern in Sample B was superimposed on another
one having a substantially larger period and a much weaker modulation, discernible in the
traces at smaller bias in the upper left inset of Fig.\;\ref{IVs-largest-Vg-effect}.
This behavior can be due to the emergence of an additional (very small) island
neighboring in line with the actual island.
Thus, in spite of a rather high homogeneity in thickness and width,
the entire nanowire can be considered as a circuit including several weaker sections with
a local increase in QPS energies per unit length, $\epsilon_{\textrm{S}}(x)$.
In the spirit of the single charge interferometer operation, Eqs.\,(\ref{Vc-mod}-\ref{Vm-ampl}),
the contributions of these weak sections to the total QPS energy should be
summed up, taking into account phases proportional to the charges induced by the gate on all intermediate islands.
The resulting quasiperiodic dependencies can be also interpreted in terms of the Aharonov-Casher effect
demonstrated in the experiments with phase-biased arrays of small JJs
(i.e. in a lumped-element analog of superconducting nanowire) \cite{Manucharyan2010,Pop2011}.
Recent calculations by Vanevi\'{c} and Nazarov \cite{Vanevic} support the hypothesis that
apparently homogeneous nanowires may naturally have a strongly inhomogeneous distribution
of specific QPS energy $\epsilon_{\textrm{S}}(x)$ because of the exponential dependence
on the local parameters. Large spatial fluctuations of the local energy gap detected by
scanning tunneling methods were also reported for thin disordered films of TiN
\cite{Sacepe2008}, NbN \cite{Mondal2011} and InO \cite{Sacepe2011}.

In conclusion, we have demonstrated the single-charge effect in superconducting
nanowires having a transistor configuration with a capacitively coupled gate,
embedded in a high-impedance environment. A deeper understanding and better control of the nanowire
parameters determining the characteristics of these transistors and other possible
circuits is urgently needed and motivates us to conduct further research.
Generally, the demonstrated duality of the QPS transistor and the dc SQUID may open the way towards
interesting applications of QPS nanocircuits in electronics and metrology.

The authors acknowledge assistance from Thomas Weimann and Peter Hinze with the
fabrication of the samples, Thomas Scheller for performing NbSi depositions and EDX measurements,
and Thorsten Dziomba for AFM measurements.
This work was partially supported by the EU through the REUNIAM
and SCOPE projects.

\end{document}